# Simultaneous Left Atrium Anatomy and Scar Segmentations via Deep Learning in Multiview Information with Attention

Guang Yang [1,2,†], Jun Chen [9], Zhifan Gao [4], Shuo Li [4], Hao Ni [5,6], Elsa Angelini [7], Tom Wong [1,2], Raad Mohiaddin [1,2], Eva Nyktari [1], Ricardo Wage [1], Lei Xu [8], Yanping Zhang [3], Xiuquan Du [3], Heye Zhang [9,†], David Firmin [1,2,‡], Jennifer Keegan [1,2,‡]

[1] Cardiovascular Research Centre, Royal Brompton Hospital, SW3 6NP, London, U.K.

[2] National Heart and Lung Institute, Imperial College London, London, SW7 2AZ, U.K.

[3] Anhui University, Hefei, 230601, China.

[4] Department of Medical Imaging, Western University, London, ON, N6A 3K7, Canada.

[5] Department of Mathematics, University College London, London, WC1E 6BT, U.K.

[6] Alan Turing Institute, London, NW1 2DB, U.K.

[7] NIHR Imperial Biomedical Research Centre, ITMAT Data Science Group, Imperial College London, London, SW7 2AZ, U.K.

[8] Department of Radiology, Beijing Anzhen Hospital, Capital Medical University, Beijing, China.

[9] School of Biomedical Engineering, Sun Yat-Sen University, Shenzhen, 510006, China.

[†] Corresponding authors: Guang Yang and Heye Zhang

Emails: g.yang@imperial.ac.uk and zhangheye@mail.sysu.edu.cn

Phone: (0044) 020 7352 8121 / (0086) 020 8327 1669; Fax: (0044) 020 7351 8699 / (0086) 020 8327 1668

Address: Cardiovascular Research Centre, Royal Brompton Hospital, SW3 6NP, London, U.K. and School of Biomedical Engineering, Sun Yat-Sen University, Shenzhen, 510006, China.

[‡] Co-last authors

# ABSTRACT

Three-dimensional late gadolinium enhanced (LGE) cardiac MR (CMR) of left atrial scar in patients with atrial fibrillation (AF) has recently emerged as a promising technique to stratify patients, to guide ablation therapy and to predict treatment success. This requires a segmentation of the high intensity scar tissue and also a segmentation of the left atrium (LA) anatomy, the latter usually being derived from a separate bright-blood acquisition. Performing both segmentations automatically from a single 3D LGE CMR acquisition would eliminate the need for an additional acquisition and avoid subsequent registration issues. In this paper, we propose a joint segmentation method based on multiview two-task (MVTT) recursive attention model working directly on 3D LGE CMR images to segment the LA (and proximal pulmonary veins) and to delineate the scar on the same dataset. Using our MVTT recursive attention model, both the LA anatomy and scar can be segmented accurately (mean Dice score of 93% for the LA anatomy and 87% for the scar segmentations) and efficiently (~0.27 seconds to simultaneously segment the LA anatomy and scars directly from the 3D LGE CMR dataset with 60–68 2D slices). Compared to conventional unsupervised learning and other state-of-the-art deep learning based methods, the proposed MVTT model achieved excellent results, leading to an automatic generation of a patient-specific anatomical model combined with scar segmentation for patients in AF.

# INTRODUCTION

Three-dimensional late gadolinium enhanced (LGE) cardiac MR (CMR) of left atrial (LA) scars in patients with atrial fibrillation (AF) has recently emerged as a promising technique to stratify patients, to guide ablation therapy and to predict treatment success [1][2][3]. Visualisation and quantification of LA scar tissue from LGE CMR require a segmentation of the LA anatomy (including proximal pulmonary veins (PV)) and a segmentation of the LA scars [4]. In clinical practice, the LA anatomy and LA scars are generally segmented by radiologists with manual operations, which are time-consuming, subjective and lack reproducibility [5]. Therefore, automatic LA anatomy and LA scar segmentation methods are highly in demand for improving the clinical workflow.

Automatic segmentations of LA anatomy and LA scar from LGE CMR are very challenging tasks due to the low visibility of the LA boundaries and the small discrete regions of the LA scars. The LGE CMR technology is widely used to visualise scar tissues by enhancing their signal intensities, while the nulling of signals from healthy tissue reduces the visibility of the LA boundaries [6]. Moreover, LA scars occupy only a small portion of the LA wall, and they distribute discretely; therefore, detection of LA scars is highly susceptible to noise interferences. In the AF patient population, prolonged scanning time, irregular breathing pattern and heart rate variability during the scan can result in poor image quality that can also further complicate both segmentation tasks. Because of these issues, previous studies have segmented the LA anatomy from an additional bright-blood data acquisition, and have then registered the segmented LA anatomy to the LGE CMR data for visualisation and delineation of the LA scars [7][8][9]. This approach is complicated by motion (bulk, respiratory or cardiac) between the two acquisitions and subsequent registration errors. Furthermore, it is based on a two-phase framework. It is inadequate to achieve accurate and efficient estimation for the LA scars because the LA anatomy and LA scars segmentations are separately handled, and no feedback connection exists between them during the algorithm training.

To address the above problems, we propose a fully automated multiview two-task (MVTT) recursive attention model to segment LA anatomy and LA scars from LGE CMR simultaneously. In the same way that reporting clinicians typically step through 2D axial slices to find correlated information while also using complementary information from orthogonal views, we parse 3D LGE CMR images into continuous 2D slices and apply 2D convolutions instead of a 3D convolution. Our proposed MVTT method mainly consists of a multiview learning network and a dilated attention network. The multiview learning network learns the correlation between 2D axial slices by a sequential learning subnetwork. At the same time, two dilated residual subnetworks learn the complementary information from the sagittal and coronal views. Then, we integrate the two kinds of complementary information into the axial slice features to obtain fused multiview features to achieve the segmentation of the LA anatomy. Since LA scars are very small, the dilated attention network learns an attention map from the image to force our network to focus on these small regions and to reduce the influence of background noise. In our proposed MVTT, the LA anatomy and LA scars share the multiview features to handle the two segmentation tasks, thus can mitigate the error accumulation problem.

Major contributions of this article are as follows:

- An MVTT framework is proposed to provide clinicians with the segmented LA anatomy and LA scars directly and simultaneously from the LGE CMR images, avoiding the need for an additional data acquisition for anatomical segmentation and subsequent registration errors.
- A multiview learning network is presented to fuse multiview features. It mainly correlates the 2D axial slices while integrating complementary information from orthogonal views to relieve the loss of 3D spatial information.
- A dilated attention model is presented to force our network to focus on the small targets of LA scars. It mainly learns an attention map for the localisation and representation of the LA scars but neglects high intensity signals from noise.

## RELATED WORK

**Segmentation of the LA Anatomy**

The LA anatomy would ideally be segmented from the cardiac and respiratory-gated LGE CMR dataset that is used to segment the scar tissue. However, this is difficult as the nulling of signal from healthy tissue reduces the visibility of the LA wall boundaries. Other options are to segment the anatomy from a separately acquired breath-hold magnetic resonance angiogram (MRA) study [8][10] or from a respiratory and cardiac gated 3D balanced steady state free precession (b-SSFP) acquisition [4][7][9]. While MRA shows the LA and PV with high contrast, these acquisitions are generally un-gated and usually acquired in an inspiratory breath-hold. The anatomy extracted from MRA can therefore be highly deformed compared to that in the LGE CMR study. Although a 3D b-SSFP acquisition takes longer to acquire, it is in the same respiratory phase as the LGE CMR and the extracted anatomy can be better matched.

Ravanelli et al. [10] manually segmented the LA wall and PV from MRA images, for which both efficiency and accuracy have been achieved. The segmented LA and PV were then mapped to the 3D LGE CMR dataset and this was followed by a thresholding based segmentation of the LA scars. Recently, Tao et al. [8] combined atlas based segmentation of LGE CMR and MRA to define the cardiac anatomy. After image fusion of the LGE CMR and MRA, accurate LA chamber and PV segmentation was achieved by a level set based local refinement, based on which an objective LA scars assessment is envisaged in future development.

Instead of using MRA, Karim et al. [7] used a respiratory and cardiac gated 3D b-SSFP acquisition to define the cardiac anatomy. This was resolved using a statistical shape model, and the LA scars were then segmented using a graph-cut model assuming that the LA wall is ~3mm from the endocardial border obtained from the LA geometry extraction. In Yang et al. [4][9] LA anatomy was derived by a whole heart segmentation (WHS) method [11] applied to the 3D b-SSFP data, and was then propagated to the corresponding LGE CMR images. All of these methods, which rely on a second bright-blood dataset (either MRA or 3D b-SSFP), are complicated by motion (bulk, respiratory or cardiac) between the two acquisitions and suffer from subsequent registration errors.

More recently, convolutional neural networks (CNN) based approaches have been proposed to segment the LA and PV [12] [13][14][15][16] and a grand challenge has been held for LA anatomy segmentation [17]. These research studies on LA anatomy segmentation can potentially be useful for LA scars segmentation although to the best of our knowledge, this has not been done to date.

**Segmentation of the LA Scar**

For segmentation of scar tissue within the LA, Oakes et al. [18] analysed the intensity histogram within the manually segmented LA wall to determine a thresh-hold above mean blood pool intensity for each slice within the 3D LGE volume. In an alternative approach, Perry et al. [19] applied k-means clustering to quantitatively assess normal and scarred tissue from manual LA wall segmentation. A grand challenge was carried out for the evaluation and benchmarking of various LA scars segmentation methods, including histogram analysis, simple and advanced thresholding, k-means clustering, and graph-cuts [1]. Although these pioneering studies have shown promising results on the segmentation and quantification of LA scars using LGE CMR images, most have relied on manual segmentation of the LA wall and PV from a second dataset (MRA or 3D b-SSFP). This has several drawbacks: (1) it is a time-consuming task; (2) there are intra- and inter-observer variations; (3) it is less reproducible for a multi-centre and multi-scanner study; and (4) there are registration errors between the LA and PV segmentation from a second dataset and the LGE CMR acquisition. Inaccurate segmentation of the LA wall and PV can further complicate the delineation of the LA scars and its quantification can be error-prone. This is potentially one of the reasons that there are currently on-going concerns regarding the correlation between LA scars identified by LGE CMR (enhanced regions) and electro-anatomical mapping systems (low voltage regions) [20][21] used during an electrophysiology procedure. Yang et al. [4][22] proposed a supervised learning based method (using Support Vector Machine or Autoencoder) to delineate LGE regions that were initially over-segmented into super-pixel patches. Although this method achieved high accuracy in LA scars segmentation fully automatically, the scar boundaries and continuity of the LA scars in 3D could be affected due to this 2D slice by slice processing. As for the LA anatomy delineation, deep learning based architectures, e.g., U-Net [23] and V-Net [24], have been proposed to solve semantic segmentation for many computer vision and medical image analysis problems including segmentation of the LA anatomy; however, to the best of our knowledge, they have not yet been developed and validated for LA scars segmentation.

## PROPOSED METHODS

Our work mimics the inspection process of radiologists who step through 2D axial slices to find correlated information while also using complementary information from orthogonal views. Hence, we slice the 3D LGE CMR volume into contiguous 2D slices and perform 2D slice segmentation. This has two major advantages: 1) it increases training data samples and 2) 2D convolution has better memory efficiency. The workflow of our MVTT is summarised as shown in Figure 1. It consists of three major subnetworks—a multiview learning network, a dilated residual network and a dilated attention network—that perform the segmentations of the LA and proximal PV and LA scars automatically and simultaneously.

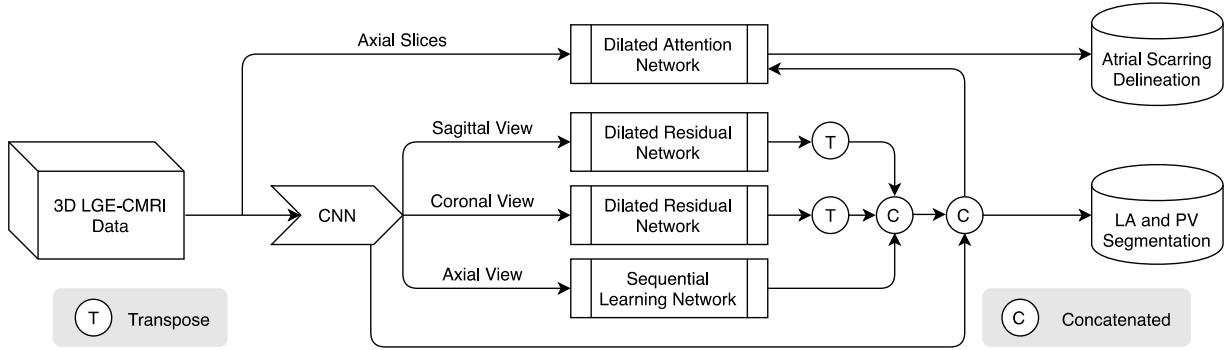

Figure 1: Overall workflow of our proposed MVTT recursive attention model that consists of three major subnetworks.

**Multiview Learning Network for Feature Fusion**

We slice the 3D LGE CMR volume into many 2D slices, thus losing the spatial correlation between these 2D slices. To learn the correlation between 2D slices in a 3D LGE CMR dataset, we investigate a sequence learning subnetwork $S(\theta_a): f_a \rightarrow F_a$, where the $\theta_a$ denotes the parameters of $S$, $f_a$ is the high-resolution feature extracted from axial 2D slices. $F_a$ represents the correlated sequence features of 2D axial slices mapped from $f_a$. The correlated sequence features are mainly learned by the convolutional long-short term memory (ConvLSTM), which is a special recursive neural network architecture that can be defined mathematically as

$$f_t = \sigma(W_{xf} * x_t + W_{hf} * h_{t-1} + W_{cf} \circ c_{t-1} + b_f) \quad (1)$$

$$i_t = \sigma(W_{xi} * x_t + W_{hi} * h_{t-1} + W_{ci} \circ c_{t-1} + b_i) \quad (2)$$

$$c_t = f_t \circ c_{t-1} + i_t \circ \text{ReLU}(W_{xc} * x_t + W_{hc} * h_{t-1} + b_c) \qquad (3)$$

$$o_t = \sigma(W_{xo} * x_t + W_{ho} * h_{t-1} + W_{cfo} \circ c_t + b_o) \qquad (4)$$

$$h_t = o_t \circ \text{ReLU}(c_t) \qquad (5)$$

where $*$ represents convolutional operator and $\circ$ denotes the Hadamard product, $W$ terms denote weight matrices, b terms denote bias vectors, $\sigma$ represents a sigmoid function and ReLU is used in our study instead of tanh. The ConvLSTM uses three gates including the input gate $i_t$, the forget gate $f_t$ and the output gate $o_t$. The memory cell $c_t$ represents an accumulator of the state information and $h_t$ denotes the hidden states.

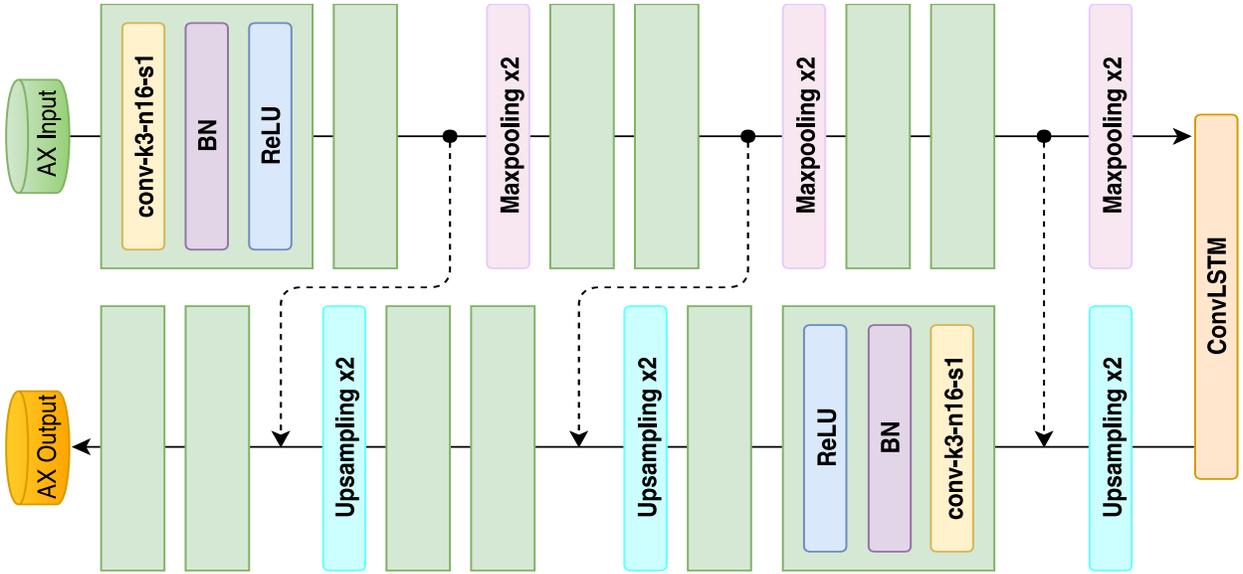

Figure 2: Architecture of the proposed sequential learning network with corresponding kernel size (k), number of feature maps (n) and stride (s) indicated for each convolutional layer.

In order to learn the complementary information from the sagittal and coronal views, we propose to use a full CNN with shortcut connections that is similar to the residual network [25]. To reduce the information loss, we introduce the dilated convolution [26], which can increase the receptive field while keeping the size of the feature map unchanged efficiently. In addition, it can aggregate multiscale contextual information with the same number of parameters. However, standard dilated convolution can cause a gridding problem. We alleviate the gridding problem by introducing a hybrid dilated convolution (HDC) into our network [27]. Thus, the complementary information from the sagittal and coronal views is learned by two dilated residual subnetworks: $S(\theta_s):f_s\to F_s$ and $S(\theta_c):f_c\to F_c$, where the $\theta$ denotes the parameters of S, $f_s$ and $f_c$ are the high-resolution features extracted from sagittal and coronal 2D slices. $F_s$ and $F_c$ represent the learned complementary information.

To compensate for the loss of spatial information in the axial view slice, we incorporate the complementary information into the correlated sequence features.

$$F_v = F_a + T(F_c) + T(F_s) \tag{6}$$

where the $T(\cdot)$ denotes the transposition operation that transposes 2D slice features of the sagittal and coronal views to the slice features of axial view. $F_v$ represents the fused features.

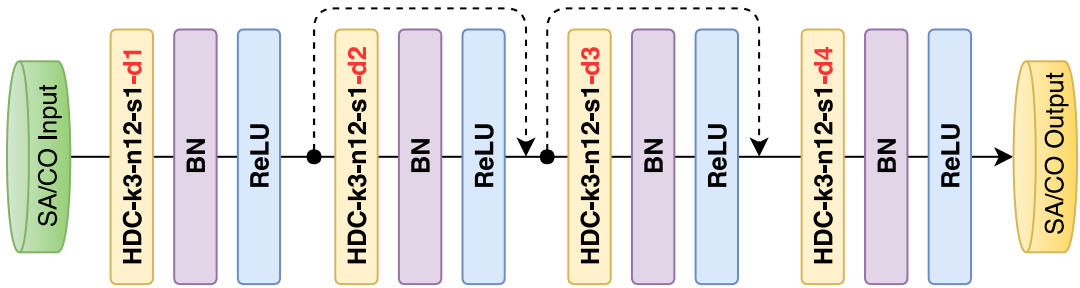

Figure 3: Architecture of the proposed dilated residual network with corresponding kernel size (k), number of feature maps (n), stride (s) and dilation rate (d) indicated for each convolutional layer.

**Dilated Attention Network for LA Scars Representation Enhancement**

Regions of LA scar are relatively small and discrete; therefore, in this study we tackle the delineation of LA scars using the attention mechanism to force the model to focus on the locations of the LA scar, and to enhance the representations of the LA scars at those locations. Conventional pooling operations can easily lose the information of these small LA scar regions. Therefore, a novel dilated attention network is designed to integrate a feed-forward attention structure with the dilated convolution to preserve the fine information of the LA scars [28]. The dilated attention network mainly learns an attention mask $S(\theta_{am}):\text{I}\to\text{AM}$, where the $\theta_{am}$ denotes the parameters of S, I is the 2D axial slice, and AM is the attention mask. In our proposed dilated attention network, the attention is provided by a mask branch, which is changing adaptively according to the learned trunk branch. We utilise a sigmoid layer, which connects to a $1\times 1$ convolutional layer to normalise the feature maps from mask branch into a range of $[0,1]$ for each channel (c) and spatial position (i) of the feature vector $x_{i,c}$ to get the AM across all the channels [28]. This sigmoid layer can be defined as following:

$$AM(x_{i,c}) = \frac{1}{1 + e^{(-x_{i,c})}} \tag{7}$$

The attention mask obtained from the mask branch is directly applied to the maps derived from the trunk branch in order to get the attention feature maps via a product operation. Because the attention mask can potentially affect the performance of the trunk branch, a skip connection with sum operation is also applied to mitigate such influence. The output O of the attention model can be denoted as

$$O(x_{i,c}) = (1 + AM(x_{i,c})) \cdot F(x_{i,c}) \tag{8}$$

in which i ranges over all spatial positions, c ranges over all the channels, $AM(x_{i,c})$ is the attention mask, which ranges from [0,1], $F(x_{i,c})$ represents the fused multiview features, and · denotes the dot product.

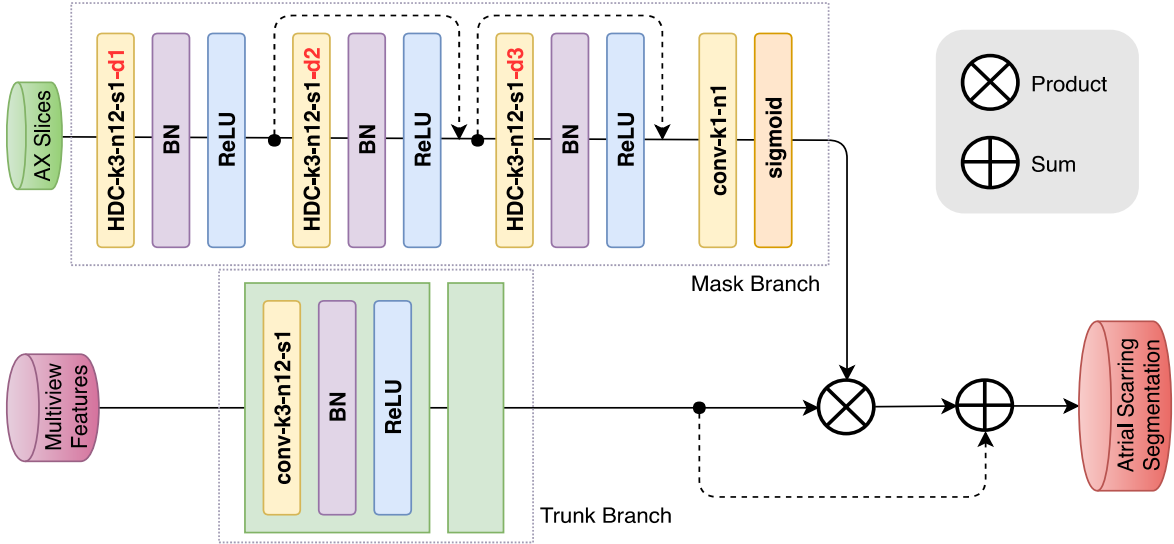

Figure 4: Architecture of the proposed dilated attention network with corresponding kernel size (k), number of feature maps (n), stride (s) and dilation rate (d) indicated for each convolutional layer.

**Hybrid Loss for Two Segmentation Task Learning**

To achieve the two segmentation tasks of delineating LA anatomy and LA scars simultaneously, our proposed MVTT shares the fused feature $F_v$. For the segmentation of LA anatomy, two convolutional layers with parameters of $\theta_l$ are used to further learn the final segmentation map of the LA anatomy. Therefore, through integrating the multiview learning network, the segmentation of LA anatomy can be achieved by the maximum likelihood estimation based on the conditional probability distributions $p(m_l/\text{I}; \theta_1)$:

$$G = \arg\max_{\theta_1} p(m_l/\text{I}; \theta_1) \tag{9}$$

where $\theta_1 = \theta_s, \theta_c, \theta_a, \theta_l$, the $m_l$ represents the probability map of the LA anatomy.

For the segmentation of LA scar, three convolutional layers with the parameters of $\theta_{as}$ are connected to further learn the segmentation map of LA scars from $F_v$ and I. Finally, the segmentation of LA scars can be achieved by the maximum likelihood estimation based on the conditional probability distributions $p(m_{as}/\text{I}; \theta_2)$:

$$G = \arg\max_{\theta_{am}, \theta_{as}} p(m_{as}/\text{I}; \theta_2) \tag{10}$$

where $\theta_2 = \theta_s, \theta_c, \theta_a, \theta_{am}, \theta_{as}$, the $m_{as}$ represents the probability map of LA scars. It is of note that there is a significant class imbalance between LA scars and background voxels. This can cause the network to pay more attention to the majority of background voxels, but neglect LA scars during training, which can lead to sub-optimal performance. In order to mitigate the class-imbalance problem, we adopt a Dice loss function to make the network biased towards the LA scars as well as the LA anatomy [29]. Hence, we use a hybrid loss:

$$L = \delta(m_l, g_l) + \delta(m_{as}, g_{as}) \tag{11}$$

where the $g_l$ and $g_{as}$ represent the ground truth of LA anatomy and LA scars respectively, and $\delta$ denotes the Dice loss function.

**Network Configuration**

Our proposed MVTT mainly consists of a multiview learning network and a dilated attention network. The multiview learning network contains three subnetworks: a sequence learning network and two dilated residual networks. The detailed configurations of the sequence learning network and the dilated residual networks are shown in Figure 2 and Figure 3. The multiview learning network mainly learns the multiview features. Based on the learned multiview features, three convolutional layers are connected to perform the segmentation of LA anatomy. First two layers contain 16 kernels with the size of 3×3 and each is followed by a BN layer and a ReLU layer. The output maps of the two layers are concatenated to connect with the last layer, which is a 3×3 convolution with one kernel and is followed by a sigmoid activation function. The detailed configuration of dilated attention network is shown in Figure 4. The dilated attention network mainly learns an (or the) enhanced feature map for LA scars. Based on the learned enhanced feature map, three convolutional layers are connected to perform the segmentation of LA scars. The first two layers contain 16 kernels with the size of 3×3 and each is followed by a BN layer and a ReLU layer. The output maps of the two layers are concatenated to connect with the last layer, which is a 1×1 convolution with one kernel and is followed by a sigmoid activation function.

# EXPERIMENTS AND RESULTS

**Data Description**

CMR data were acquired in patients with longstanding persistent AF on a Siemens Magnetom Avanto 1.5T scanner (Siemens Medical Systems, Erlangen, Germany). Transverse navigator-gated 3D LGE CMR [18][30] was performed using an inversion prepared segmented gradient echo sequence (TE/TR 2.2ms/5.2ms) 15 minutes after gadolinium administration (Gadovist—gadobutrol, 0.1mmol/kg body weight, Bayer-Schering, Berlin, Germany) [31]. The inversion time (TI) was set to null the signal from normal myocardium and varied on a beat-by-beat basis, dependent on the cardiac cycle length [6]. Detailed scanning parameters are: 30–34 slices at $(1.4–1.5)\times(1.4–1.5)\times 4 mm^3$, reconstructed to 60–68 slices at $(0.7–0.75)\times(0.7–0.75)\times 2mm^3$, field-of-view $380\times 380 mm^2$. For each patient, prior to contrast agent administration, coronal navigator-gated 3D b-SSFP (TE/TR 1ms/2.3ms) data were acquired with the following parameters: 72–80 slices at $(1.6–1.8)\times(1.6–1.8)\times 3.2 mm^3$, reconstructed to 144–160 slices at $(0.8–0.9)\times(0.8–0.9)\times 1.6mm^3$, field-of-view $380\times 380\ mm^2$. Both LGE CMR and b-SSFP data were acquired during free-breathing using a prospective crossed-pairs navigator positioned over the dome of the right hemi-diaphragm with navigator acceptance window size of 5mm and CLAWS respiratory motion control [31][32]. Navigator artefact resulting from the use of a navigator restore pulse in the LGE acquisition was reduced by introducing a navigator-restore delay of 100 ms [32]. In agreement with the local regional ethics committee, CMR data were collected from 2011–2018 for persistent AF patients. The image quality of each 3D LGE dataset was scored by a senior cardiac MRI physicist on a Likert-type scale—0 (non-diagnostic), 1 (poor), 2 (fair), 3 (good) and 4 (very good)—depending on the level of SNR, appropriate TI, and interference from navigator and/or other artefact. In total, 190 cases (out of a total of 202) with image quality greater or equal to 2 were retrospectively entered into this study. This included 97 pre-ablation cases (~93% of all) and 93 post-ablation scans cases (~95% of all). Manual segmentations of the LA anatomy and LA scars were performed by a cardiac MRI physicist with >3 years of experience and specialised in LGE CMR with consensus from a second senior radiologist (>25 years of experience and specialised in cardiac MRI), which were then used as the ground truth for training and evaluation of our MVTT recursive attention model.

**Evaluation Metrics**

The evaluation has been done quantitatively using multiple metrics, e.g., the Dice score and also the segmentation accuracy, sensitivity and specificity considering that the semantic segmentation is essentially solving a classification problem [4][33]. In addition, for the LA scars segmentation, we also calculate the correlation between the LA scars extent [4] derived from the segmentation algorithms and the ground truth by assuming the LA wall thickness is fixed at 2.25mm [34].

**Implementation Details**

We used the Adam method to perform the optimisation with a decayed learning rate (the initial learning rate was 0.001 and the decay rate was 0.98). Our deep learning model was implemented using Tensorflow 1.2.1 on an Ubuntu 16.04 machine, and was trained and tested on an NVidia Tesla P100 GPU (3584 cores and 16GB GPU memory).

Training multiple subnetworks with limited data may pose a risk of over-fitting. In this study, we applied two strategies to mitigate the issue. First, we applied the early stopping strategy, which can be considered as an additional and efficient regularisation technique to avoid over-fitting. Second, we used networks with a moderate number of parameters for each subnetwork in our framework to find a balance between a sufficient complexity to perform an accurate segmentation and a relatively low likelihood of over-fitting.

In order to test the efficacy of our proposed MVTT recursive attention model, we retrospectively studied 190 3D LGE CMR scans, and divided these data into a training/ten-fold cross-validation dataset (170 3D scans) and an independent testing dataset (20 3D scans with randomly selected 10 pre-ablation and 10 post-ablation cases). For the ten-fold cross-validation, we divided the 170 scans into 10 folds randomly. Each fold contains 17 scans. When training the model, 153 scans were used as training data and the remaining 17 scans were used for testing. We performed the cross-validation loop ten times to test the stability of our proposed methods.

We pre-processed the data with the mean normalisation:

$$I' = \frac{I - \mathrm{mean}(I)}{\mathrm{max}(I) - \mathrm{min}(I)} \quad (12)$$

where *I* represents the voxel intensities of the image. It is worth noting that we performed the mean normalisation on each slice of the 3D image instead of using the entire 3D image.

Table 1: Quantitative results (mean±standard deviation) of the cross-validated LA and PV segmentation, compared to the performance using the WHS, 2D U-Net, 3D U-Net, 2D V-Net and 3D V-Net. AC: Accuracy, SE: Sensitivity, SP: Specificity and DI: Dice score.

| Methods | AC (%) | SE (%) | SP (%) | DI (%) |
| --- | --- | --- | --- | --- |
| WHS | 99.62±0.21 | 80.86±18.07 | 99.88±0.14 | 84.54±15.11 |
| 2D U-Net | 98.60±0.42 | 93.50±3.73 | 99.11±0.43 | 91.97±2.42 |
| 3D U-Net | 98.48±0.05 | 93.02±3.35 | 99.04±0.44 | 90.58±2.64 |
| 2D V-Net | 98.36±0.58 | 92.20±4.91 | 98.98±0.50 | 90.66±3.15 |
| 3D V-Net | 98.47±0.46 | 94.43±3.33 | 98.89±0.44 | 91.37±2.48 |
| Vesal et al. | 98.30±0.71 | 94.97±3.02 | 98.65±0.73 | 90.58±3.40 |
| SV+CLSTM | 98.49±0.40 | 92.41±4.59 | 99.17±0.45 | 91.67±3.12 |
| MV | 98.04±0.89 | 90.95±4.69 | 98.76±0.59 | 89.03±4.14 |
| S-LA/PV | 98.55±0.51 | 95.32±3.08 | 98.88±0.50 | 91.87±2.68 |
| MVTT | 98.62±0.46 | 92.92±4.47 | 99.20±0.38 | 92.11±2.39 |

**Performance of the LA Anatomy Segmentation**

The experimental results show that our MVTT framework can accurately segment the LA and PV (Table 1 and Table 2). The accuracy, sensitivity, specificity and Dice scores are 98.59%, 91.96%, 99.36% and 93.11% via independent testing (Table 2). The additive value of including the multiview learning and CLSTM is apparent from higher Dice scores. Figure 5 shows example segmentation results of the LA anatomy for example pre- and post-ablation cases from the independent testing dataset.

Table 2: As Table 1, but using the independent testing dataset.

| Methods | AC (%) | SE (%) | SP (%) | DI (%) |
| --- | --- | --- | --- | --- |
| WHS | 99.53±0.21 | 80.31±17.66 | 99.83±0.14 | 82.94±14.39 |
| 2D U-Net | 98.60±0.36 | 90.86±2.18 | 99.50±0.19 | 93.08±1.58 |
| 3D U-Net | 98.49±0.26 | 93.22±1.75 | 99.11±0.19 | 92.74±1.22 |
| 2D V-Net | 98.33±0.51 | 89.94±2.90 | 99.32±0.24 | 91.84±2.25 |
| 3D V-Net | 98.44±0.29 | 89.33±2.18 | 99.50±0.14 | 92.21±1.36 |
| Vesal et al. | 98.54±0.25 | 91.54±1.76 | 99.36±0.19 | 92.81±1.37 |
| SV+CLSTM | 98.49±0.40 | 90.07±2.51 | 99.48±0.27 | 92.54±1.69 |
| MV | 97.83±0.64 | 89.19±3.02 | 98.84±0.44 | 89.48±4.14 |
| S-LA/PV | 98.56±0.44 | 90.81±3.17 | 99.48±0.20 | 90.28±26.83 |
| MVTT | 98.59±0.40 | 91.96±2.11 | 99.36±0.28 | 93.11±1.86 |

Table 3: Quantitative results (mean±standard deviation) of the cross-validated LA scars delineation. For the LA scars delineation, we compared with SD based thresholding (2-SD), k-means, Fuzzy c-means, 2D U-Net, 3D U-Net, 2D V-Net and 3D V-Net. AC: Accuracy, SE: Sensitivity, SP: Specificity and DI: Dice score.

| Methods | AC (%) | SE (%) | SP (%) | DI (%) |
|---|---|---|---|---|
| 2-SD | 99.89±0.05 | 76.37±19.43 | 99.92±0.06 | 57.84±18.07 |
| K-means | 99.85±0.04 | 74.78±15.17 | 99.89±0.05 | 49.80±16.27 |
| Fuzzy c-means | 99.85±0.04 | 78.67±14.24 | 99.88±0.06 | 49.95±17.45 |
| 2D U-Net | 99.92±0.04 | 87.94±8.90 | 99.94±0.03 | 77.86±9.03 |
| 3D U-Net | 99.92±0.04 | 77.43±14.26 | 99.96±0.03 | 74.39±12.06 |
| 2D V-Net | 99.92±0.04 | 84.51±11.19 | 99.94±0.03 | 75.01±11.80 |
| 3D V-Net | 99.92±0.04 | 83.97±8.90 | 99.95±0.03 | 74.28±13.80 |
| Vesal et al. | 99.94±0.03 | 77.18±14.5 | 99.97±0.02 | 75.25±12.3 |
| MV+AT | 99.93±0.04 | 82.21±10.35 | 99.96±0.03 | 79.38±10.62 |
| MV+CLSTM | 99.40±0.03 | 81.09±12.58 | 99.97±0.02 | 81.12±9.79 |
| SV+CLSTM+AT | 99.93±0.04 | 82.47±10.42 | 99.96±0.02 | 78.68±9.26 |
| S-Scar | 99.95±0.03 | 80.65±9.77 | 99.97±0.01 | 82.32±8.36 |
| MVTT | 99.94±0.03 | 85.88±9.84 | 99.97±0.02 | 82.58±8.72 |

Table 4: As Table 3, but using the independent testing dataset.

| Methods | AC (%) | SE (%) | SP (%) | DI (%) |
|---|---|---|---|---|
| 2-SD | 99.84±0.06 | 81.38±19.51 | 99.87±0.07 | 51.66±19.50 |
| K-means | 99.82±0.06 | 75.92±13.76 | 99.86±0.06 | 47.78±15.24 |
| Fuzzy c-means | 99.82±0.04 | 77.78±13.76 | 99.86±0.05 | 48.09±16.26 |
| 2D U-Net | 99.93±0.03 | 89.14±4.66 | 99.95±0.01 | 80.21±9.61 |
| 3D U-Net | 99.93±0.04 | 75.22±11.31 | 99.98±0.01 | 79.41±8.02 |
| 2D V-Net | 99.92±0.03 | 86.39±6.30 | 99.95±0.02 | 79.45±8.60 |
| 3D V-Net | 99.92±0.04 | 77.07±9.72 | 99.98±0.01 | 77.05±9.96 |
| Vesal et al. | 99.92±0.03 | 83.71±10.4 | 99.96±0.03 | 76.13±10.9 |
| MV+AT | 99.93±0.04 | 70.83±9.44 | 99.99±0.01 | 79.05±8.66 |
| MV+CLSTM | 99.30±0.03 | 65.36±11.17 | 99.99±0.01 | 77.10±9.26 |
| SV+CLSTM+AT | 99.94±0.03 | 81.27±7.88 | 99.97±0.01 | 82.36±6.51 |
| S-Scar | 99.94±0.03 | 72.39±8.88 | 99.99±0.01 | 80.22±0.01 |
| MVTT | 99.95±0.02 | 86.77±4.64 | 99.98±0.01 | 86.59±5.60 |

**Performance of the LA Scars Segmentation**

Our MVTT framework has also performed well for segmenting the LA scars (Table 3 and Table 4). We achieve an overall scar segmentation accuracy of 99.95%, with a sensitivity of 86.77%, a specificity of 99.98% and a Dice score of 86.59% on the independent testing dataset (Table 4). The additive value of multiview learning, CLSTM and attention mechanism is seen through higher dice scores. Figure 6 shows LA scars segmentations from all methods in an example pre- and post-ablation patient. Visualization of the atrial scar segmentation using our MVTT (Figure 6 (b) and (k)) shows excellent agreement with the ground truth (Figure 6 (a) and (j)). In addition, Figure 7 shows the 3D segmentation results of the LA anatomy overlaid with LA scars showing high consistency compared to the manual delineated ground truth.

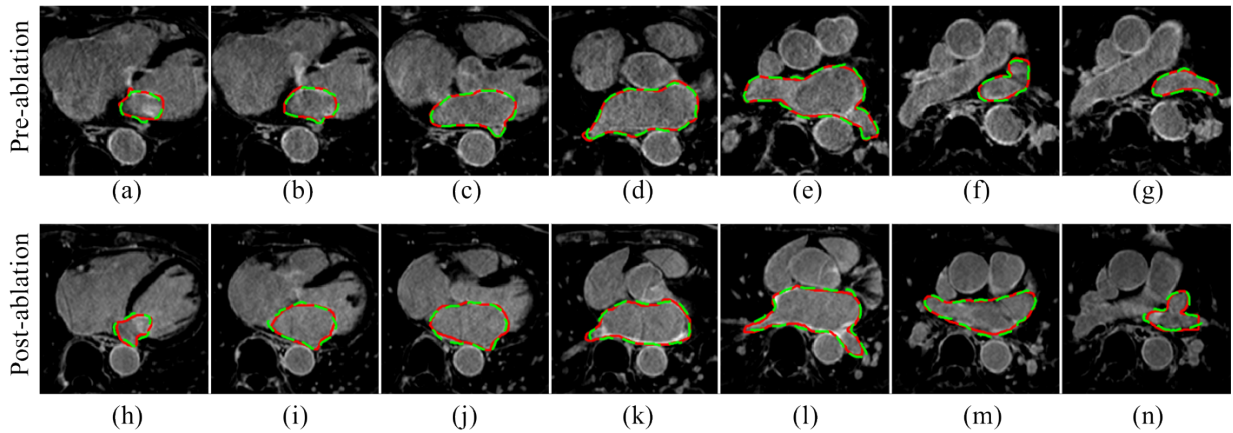

Figure 5: Qualitative visualisation of the LA anatomy segmentations (via independent testing) in multiple slices from an example pre-ablation (a-g) and an example post-ablation (h–n) study. Red contour: manual delineated ground truth. Green contour: segmentation using MVTT.

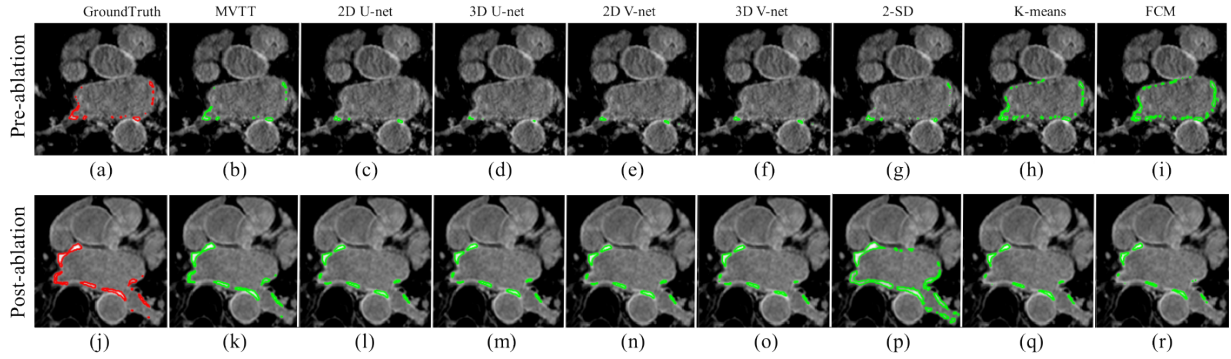

Figure 6: Qualitative visualisation of LA scars delineation (independent testing results) in an example pre-ablation (a–i) and post-ablation (j–r) study using different methods. Red = manually segmentation (ground truth), green = algorithm segmentation.

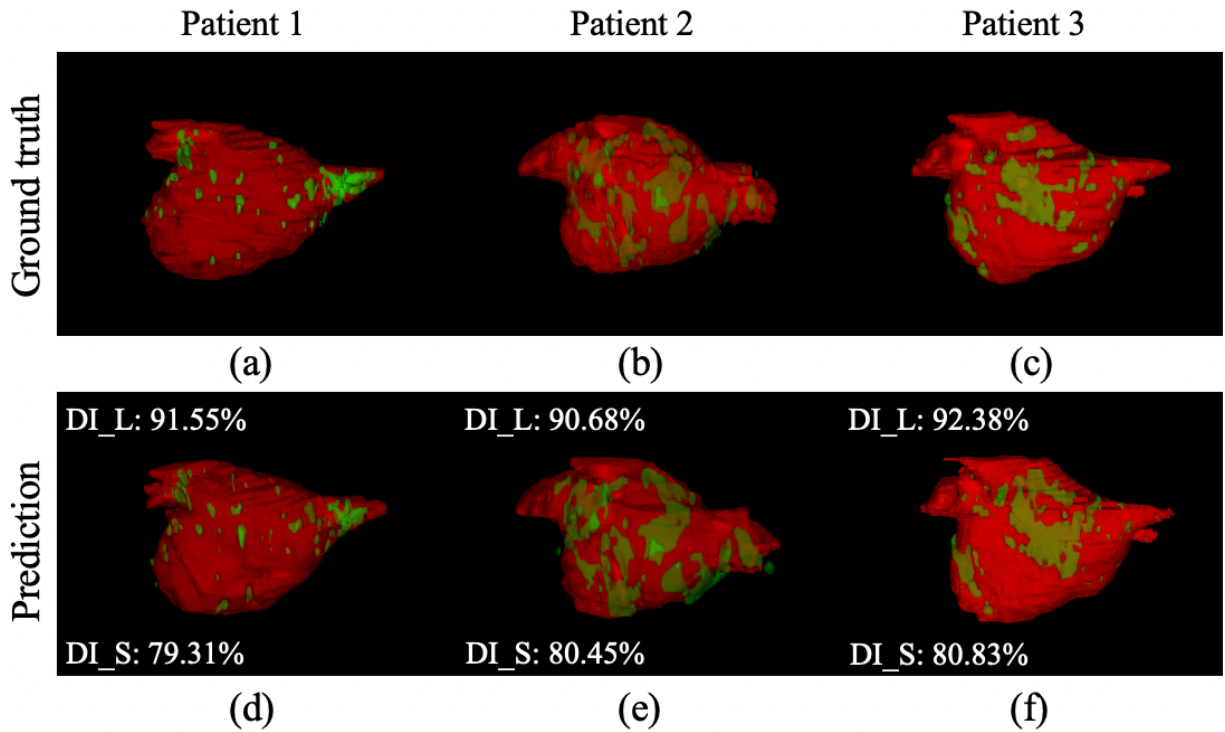

Figure 7: 3D visualization for LA anatomy and LA scars of the independent testing results (DI_L represents the DI value for predicted LA anatomy. DI_S represents the DI value for predicted LA scars). (a-c) Ground truth and (d-f) Segmentation results of using our MVTT method.

**Model Variation Studies**

To demonstrate the additive value of the multiview learning, convLSTM and attention mechanism, we performed several model variation studies: (1) For the LA and PV segmentation, we compared our MVTT model with the single axial view learning with ConvLSTM (SV+CLSTM) and multiview learning without using ConvLSTM (MV); (2) For the LA scars segmentation, we tested the multiview learning with attention network but without ConvLSTM (MV+AT), multiview learning with ConvLSTM but without attention network (MV+CLSTM), the single axial view learning with attention network and ConvLSTM (SV+CLSTM+AT). In order to prove that our MVTT was effective for delineating both LA and PV and LA scars simultaneously, we also implemented two single segmentations of LA and PV and LA scars (S-LA/PV and S-Scar).

Results on both cross-validation and independent testing showed that our MVTT model yielded superior results (see Table 1, Table 2, Table 3 and Table 4). In particular, for the LA scars delineation, our MVTT improved the Dice scores from 77%–82% to ~86% (Table 4). We also showed that our MVTT model could accurately segment LA and PV with LA scar simultaneously instead of performing these two tasks sequentially (rows S-LA/PV and S-Scar in Table 1, Table 2, Table 3 and Table 4). These superior results obtained by our proposed MVTT indicate that the effectiveness of the ConvLSTM for the sequence learning, the multiview learning for the information complement and the attention mechanism for the small target learning. In addition, it also demonstrates the effective integration of multiview learning, convLSTM and attention mechanism for the simultaneous segmentation of LA and LA scars.

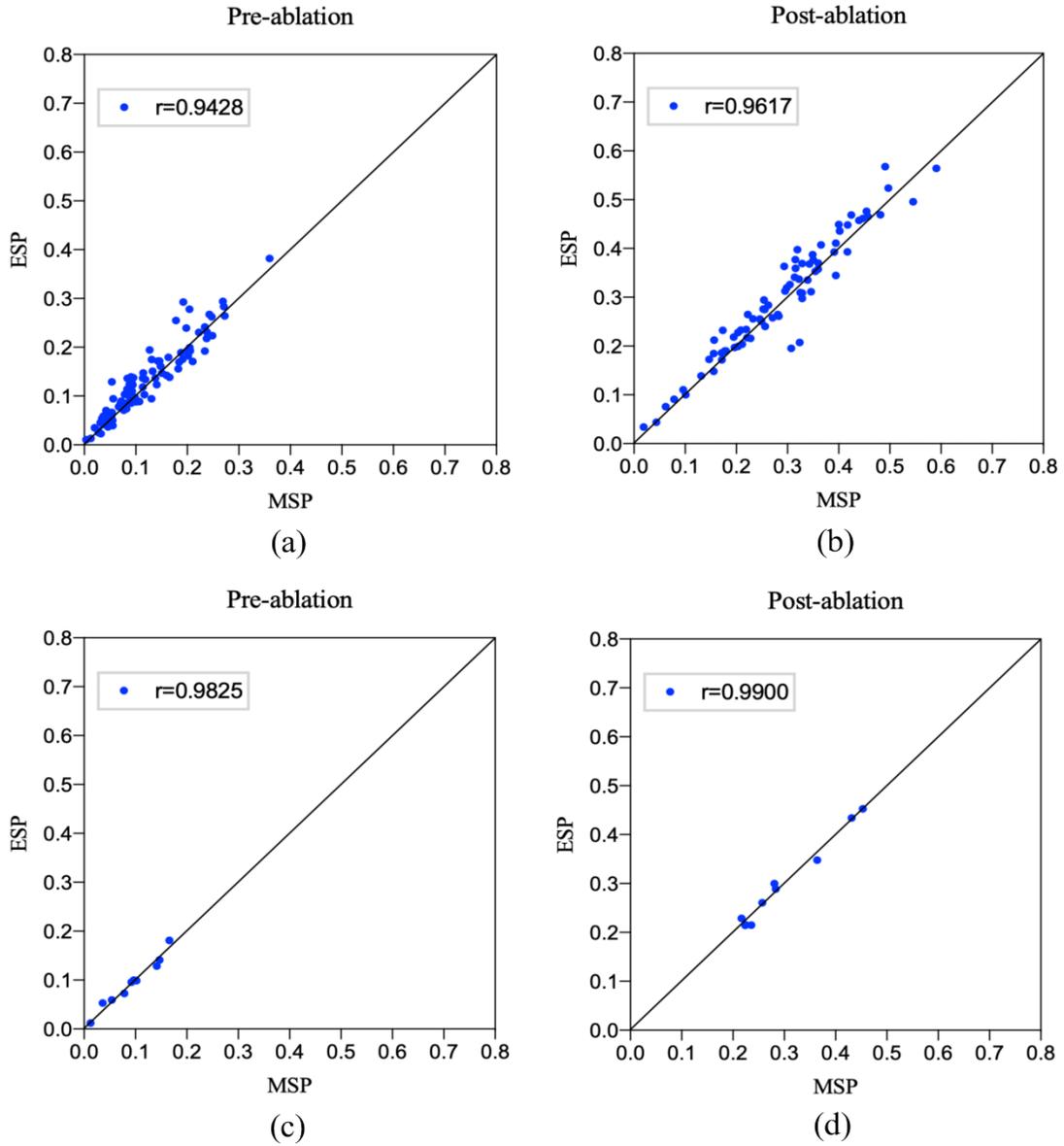

Figure 8: Correlation between the estimated LA scars percentage (ESP) of our MVTT method and the LA scars percentage from the manual delineation (MSP) (diagonal lines represent lines of identity). (a) and (b) show the correlations for pre and post ablation studies in the training/cross-validation datasets, and (c) and (d) show the correlations for pre and post ablation studies in the independent testing datasets.

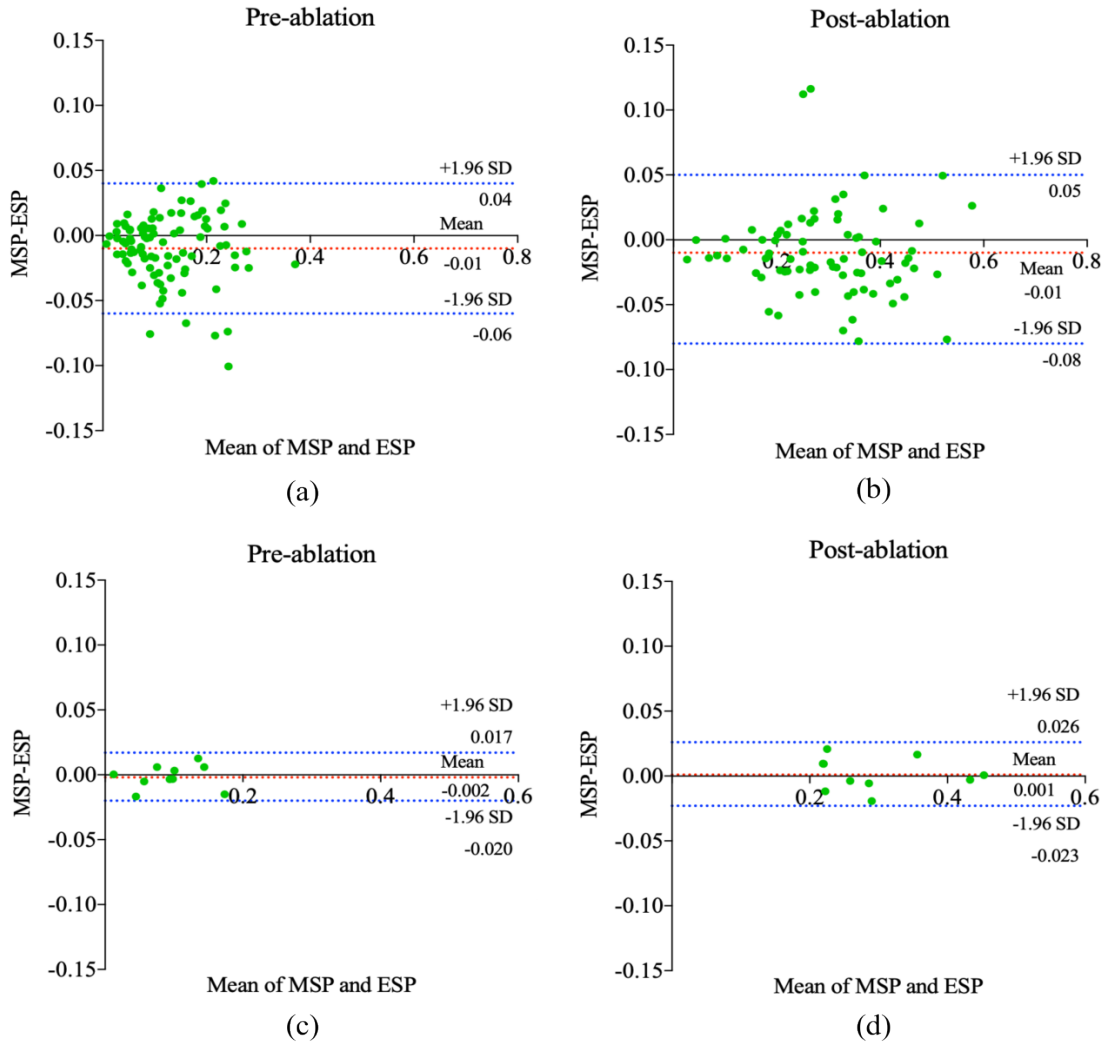

Figure 9: Bland-Altman plots for the calculated LA scars percentage of our MVTT method and the LA scars percentage of the manual delineation. (a) and (b) were calculated on the 170 LGE CMR images using training/cross-validation results. (c) and (d) were calculated on the 20 LGE CMR images using independent testing results. Horizontal lines show the mean difference and the 95% CI of limits of agreement (confidence limits of the bias), which are defined as the mean difference plus/minus 1.96 times the standard deviation of the differences. The mean differences are near the 0-line (bias=−1% [95% CI −6% to 4%] and bias=−1% [95% CI −8% to 5%] for the pre-ablation and post-ablation cases respectively via training/cross-validation and bias=−0.2% [95% CI −2% to 1.7%] and bias=−0.1% [95% CI −2.3% to 2.6%] for the pre-ablation and post-ablation cases respectively via independent testing. In summary, no significant systematic differences between the two methods can be discerned. MSP: Manual Segmented Atrial Scar Percentage; ESP: Estimated Atrial Scar Percentage.

**Model Parameter Validation**

To demonstrate the parameter effectiveness in our proposed MVTT. We carried out three extra experiments: (1) We replaced the kernel size of 3 × 3 with 5 × 5 in our proposed MVTT (K5) for kernel validation; (2) The activation function of ReLU was applied to convolutional LSTM to validate the LSTM performance (AFT); (3) For learning of dilated convolution, we replaced the dilated convolution with general convolution (NDC). The experiment results on both cross-validation and independent testing are shown in the Table 6 and Table 7. As shown in the two tables, for the validations of the kernel size, the activation function of convolutional LSTM and the dilated convolution, our proposed MVTT with the 3 × 3 kernel size, the activation function of ReLU for convolutional LSTM and the HDC can obtain the superior results. These superior results can be explained by that (1) 3 × 3 kernel size can reduce the parameters of MVTT to decrease the overfitting of model compared to the 5 × 5 kernel size; (2) ReLU can reduce the vanishing gradient problem; (3) The HDC can alleviate the gridding problem to help extract more robust features.

Table 6: Comparison of different parameter settings for the LA segmentation.

| Type | Method | AC (%) | SE (%) | SP (%) | DI (%) |
|---|---|---|---|---|---|
| 10-fold cross-validation | AFT | 98.60 ± 0.78 | 92.49 ± 6.04 | 99.23 ± 0.38 | 91.95 ± 3.88 |
| | K5 | 98.67 ± 0.47 | 94.24 ± 3.50 | 99.13 ± 0.45 | 91.47 ± 1.68 |
| | NDC | 98.61 ± 0.43 | 92.36 ± 4.17 | 99.24 ± 0.38 | 91.96 ± 2.31 |
| | MVTT | 98.62 ± 0.46 | 92.92 ± 4.47 | 99.20 ± 0.38 | 92.11 ± 2.39 |
| Independent testing | AFT | 98.41 ± 0.49 | 89.00 ± 3.66 | 99.52 ± 0.23 | 92.11 ± 2.16 |
| | K5 | 98.57 ± 0.35 | 90.95 ± 2.24 | 99.46 ± 0.26 | 92.95 ± 1.68 |
| | NDC | 98.44 ± 0.40 | 89.14 ± 2.65 | 99.53 ± 0.22 | 92.21 ± 1.75 |
| | MVTT | 98.59 ± 0.40 | 91.96 ± 2.11 | 99.36 ± 0.28 | 93.11 ± 1.86 |

Table 7: Comparison of different parameter settings for the scar segmentation.

| Type | Method | AC (%) | SE (%) | SP (%) | DI (%) |
|---|---|---|---|---|---|
| 10-fold cross-validation | AFT | 99.95 ± 0.04 | 78.83 ± 12.22 | 99.98 ± 0.02 | 81.83 ± 9.56 |
| | K5 | 99.94 ± 0.04 | 79.54 ± 11.24 | 99.97 ± 0.01 | 81.03 ± 8.13 |
| | NDC | 99.94 ± 0.03 | 77.26 ± 11.74 | 99.98 ± 0.02 | 80.07 ± 8.71 |
| | MVTT | 99.94 ± 0.03 | 85.88 ± 9.84 | 99.97 ± 0.02 | 82.58 ± 8.72 |
|  | ATF | 99.94 ± 0.03 | 75.03 ± 7.86 | 99.99 ± 0.00 | 82.80 ± 6.86 |
| Independent | K5 | 99.94 ± 0.03 | 87.87 ± 7.65 | 99.96 ± 0.01 | 82.80 ± 6.98 |
| testing | NDC | 99.94 ± 0.03 | 78.74 ± 7.81 | 99.98 ± 0.00 | 82.60 ± 7.17 |
|  | MVTT | 99.95 ± 0.02 | 86.77 ± 4.64 | 99.98 ± 0.01 | 86.59 ± 5.60 |

**Comparison Studies**

For comparison studies, we also evaluated the performance of state-of-the-art methods for LA and PV segmentation: atlas based WHS [11], Vesal et al. [35] and re-implementations of the 2D U-Net [23], 3D U-Net [36], 2D V-Net [24] and 3D V-Net [29]. For the LA scars segmentation, we compared our MVTT method with both unsupervised learning based methods [1] and also with newly developed deep learning based methods (re-implementation of Vesal et al. [35], 2D/3D U-Net and V-Net [20,21,26,31]).

**LA and PV Segmentation:** Compared to WHS, our MVTT framework obtained much higher sensitivity (91.96% vs. 80.31%) and similar specificity and therefore a higher Dice score (93.11% vs. 82.94%) in the independent testing dataset. Our MVTT model also showed better quantitative results compared to other deep learning based models (Table 1 and Table 2).

**LA Scars Segmentation:** Dice scores for pre- and post-ablation studies in the training/cross-validation and independent datasets are shown in Figure 10 and in Table 3 and Table 4. All the unsupervised learning methods, e.g., SD based thresholding and clustering, obtained high specificities, but very low sensitivities and poor Dice scores. Qualitative visualization in Figure 6 shows that the 2-SD, k-means and Fuzzy c-means (FCM) methods clearly over-estimated the enhanced scar regions, especially for the pre-ablation cases. The U-Net and V-Net based methods improved the delineation, but were still struggling to segment the LA scars accurately. Using the independent testing dataset, our MVTT model achieved a Dice score of 87% for the LA scars segmentation (83% ± 6% for the pre-ablation cases and 91% ± 3% for the post-ablation cases).

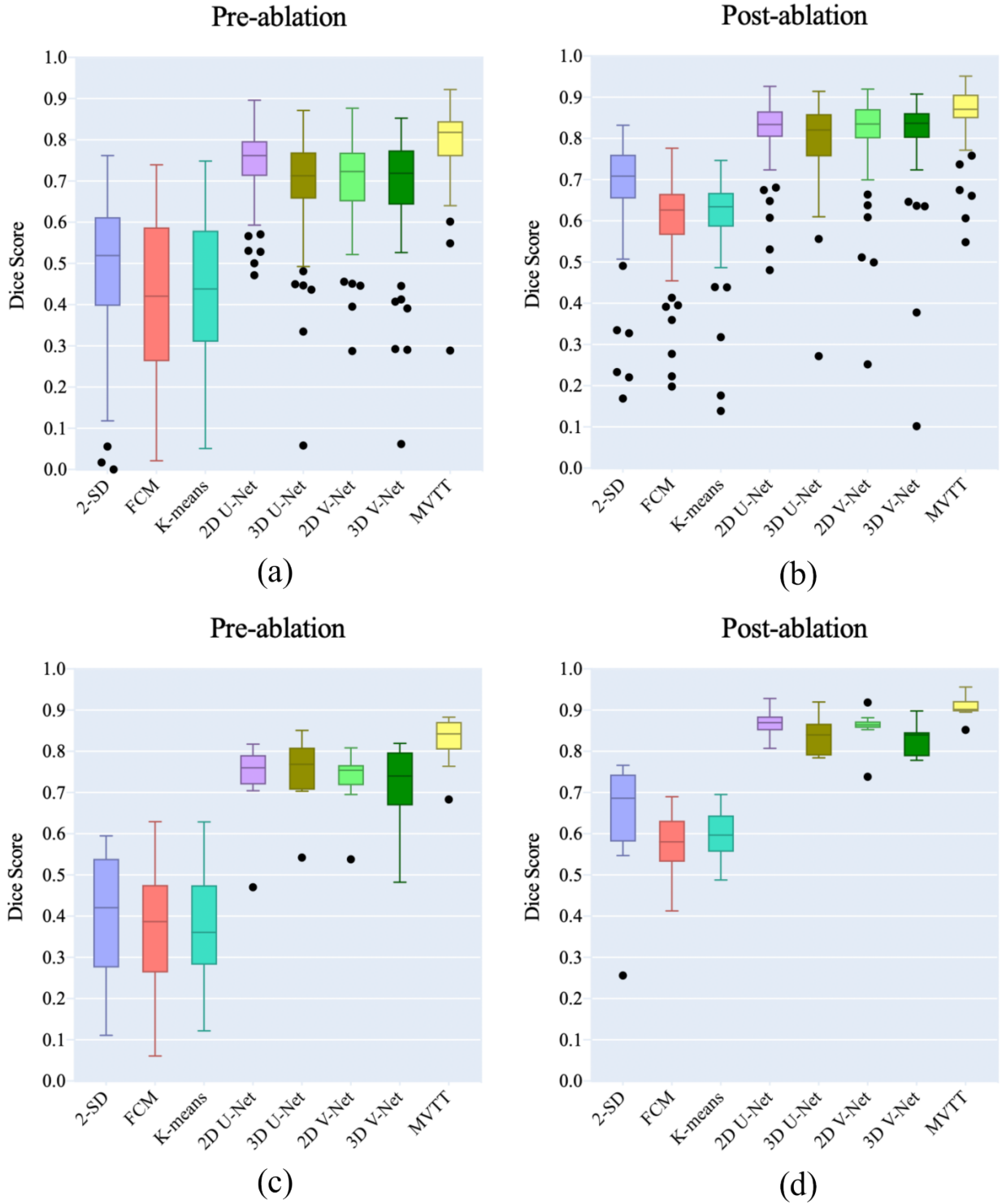

Figure 10: Boxplot of the Dice scores for comparison studies on LA scars segmentation. Training/cross-validation on the pre-ablation (a) and post-ablation (b) cases. Independent testing on the pre-ablation (c) and post-ablation (d) cases.

The superior results achieved by our proposed MVTT are mainly derived from the following aspects: (1) We fully consider the limited data, thus slicing the 3D LGE CMR volume into contiguous 2D slices to augment data. At the same time, we integrate the multiview features to improve the feature effectiveness for segmentation target learning; (2) We fully consider the small target learning for LA scars that a dilated attention mechanism is proposed to focus on small LA scars for its accurate learning. (3) We fully consider the multi-task learning that leverages the shared features to improve the segmentation performance.

**Analysis of Potential Practical Application**

We proposed an automated method to segment the LA with proximal PV and LA scars aiming to use such information to stratify AF patients, guide ablation therapy and predict treatment success. Patient stratification is based on scar burden defined as the LA scar tissue as a percentage of the LA volume. Hence, we further analyse the calculated scar percentage between our MVTT and the ground truth. Figure 8 shows the linear regression analysis of the calculated scar percentage between our MVTT and the ground truth for both training and independent testing datasets. The Pearson correlation coefficients for the independent testing data show excellent agreement between the two (r = 0.983, 95% CI 0.966 to 0.996 [pre-ablation] and r = 0.990, 95% CI 0.950 to 0.998 [post-ablation] $\in [0.8,1.0]$). Bland-Altman plots showing the difference in scar percentage (between our MVTT and manual segmentations) against the manual segmentation scar percentage (as gold standard) are presented in Figure 9. From these figures, we find that our calculated scar percentage has high consistency with manual delineation by our physicist. For the independent testing, it took 5.34 seconds to segment 20 cases (to derive both the LA anatomy with proximal PV and LA scars simultaneously), and therefore ~0.27 seconds per case, which has similar performance compared to the 2D U-Net and 2D V-Net models (~0.2 seconds per case) and faster compared to the 3D U-Net and 3D V-Net models (~1.12 seconds and ~0.46 per case). These results have indicated the potential of our proposed MVTT in real clinical applications.

## DISCUSSION

In this study, we have developed a fully automatic MVTT deep learning framework for segmenting both LA and atrial scar simultaneously. Our MVTT framework combines a sequential learning network that imitates 3D data scrutinisation routinely performed by the reporting clinicians and a dilated residual learning network and an attention model to delineate the LA scars more accurately. Our proposed framework only requires a 3D LGE CMR dataset as the input and avoids acquiring/using additional scans for the delineation of the cardiac anatomy. In addition to reducing scanning time, this also eliminates the inevitable errors which occur when multiple datasets are registered. This has been achieved mainly because (1) our 3D LGE CMR studies are reliable so that most scans (~93.27% of all pre-ablation cases and ~94.90% of all post-ablation cases) can be used for training, validation and testing and (2) our developed MVTT framework is robust to detect and segment not only the LA anatomy but also the LA scars, which are relatively small. Our segmentation results have been validated against manual ground truth delineation carried out by experienced physicists and radiologists and have demonstrated promising potential for a direct application in clinical environment.

The performance of our proposed MVTT model did not rely on a comprehensive tuning of network parameters. In our preliminary study [37], we found that our initial MVTT model suffered from over-fitting by visualising the loss functions of training/cross-validation. We subsequently incorporated an early stopping strategy that has effectively reduced this and resulted in excellent performance in the independent testing dataset. Compared with our preliminary study, a new dilated attention network and a new dilated residual network, which integrated the hybrid dilated convolution, were proposed for a more efficient feature extraction and a more efficient generation of the attention map. In addition, we replaced the mean squared loss in our preliminary study with the Dice loss to further focus on the problem of small target segmentation. Furthermore, experiments in our current work were extended to a larger database with 190 cases, and more experiment validations and detailed discussions were added for the current study.

A limitation of this work is that the 'ground truth' segmentations that our MVTT framework was developed from and validated against were derived manually. While this is not ideal due to intra and inter operator variability, it is the most commonly used method for establishing the ground truth for such tasks and there is no real alternative available. Our ground truth was determined by a single expert due to limited resources and we are unable to provide an assessment of inter-rater agreement. While this is not ideal, the single-expert delineations were checked by a second expert who made changes (by consensus) if necessary.

In addition, many studies have demonstrated that multiscale network is an efficient architecture to acquire different receptive fields and capture information at different scales to improve the performance of a trained deep learning model [38][39]. However, integrating multiscale network into our MVTT will further increase the network complexity. It requires further investigations on how to make the combination of MVTT and multiscale network work efficiently.

A key challenge of imaging LA scars using LGE CMR remains the limited spatial resolution [40]. Autopsy studies showed that the mean LA transmural thickness is 2.2–2.5mm (endocardium-epicardium) [34] but this may be reduced for persistent AF patients [41]. Most current 3D LGE CMR sequences have a spatial resolution about 1–2mm [18][30][42][43], which is usually reconstructed/interpolated to a higher value [43]; however, current LGE CMR sequences still suffer from partial volume effects [40] and this may affect the delineation and quantification of the LA scars. Furthermore, the quantification of LA scars is based on the segmentation of LA (and proximal PV) and LA scars. It is important for us to perform the segmentation and quantification of LA scars simultaneously to further improve the efficiency [44].

We have performed comprehensive comparison studies in the current work—comparing our results with conventional unsupervised learning based methods and supervised deep learning models. It is of note that the WHS method derived the LA anatomy from additionally acquired bright-blood image that was then registered to the LGE-MRI for the further scar segmentation. Our MVTT method derived both LA anatomy and scar segmentations from a single 3D LGE CMR dataset. This is a challenging task which eliminates the need for an additional acquisition to define atrial anatomy and subsequent registration errors. Interestingly, by comparing Table 1 and Table 2 with Table 3 and Table 4, we found that the U-Net and V-Net based methods achieved a Dice score over 90% for the LA and PV segmentation, but the performance of these methods was much worse for the LA scars delineation (<81% Dice score). This may be due to the fact that these U-Net and V-Net based architectures are more suitable for segmenting relatively larger areas but are not so effective on small LA scars regions.

## CONCLUSIONS

In this study, we propose a fully automatic MVTT recursive attention model, which consists of three major subnetworks that incorporate multiview learning, convLSTM and attention mechanism. The proposed MVTT model can resolve the connections in-between the axial image slices and preserve the overall information from the other two views. This intuitively mimics the way reporting clinicians scrutinise the 3D data. For the abnormal and small LA scars regions, our developed attention network also imitates the human attention mechanism that can efficiently exclude interferences and lets the network focus on the abnormalities it tries to segment. Validation has been performed against manually defined ground truth, and both model variation studies and comparison studies demonstrate the efficacy of our MVTT model in pre- and post-ablation studies. In conclusion, the proposed MVTT framework outperformed other state-of-the-art methods and it can be integrated into the clinical routine for a fast, reproducible and reliable LA scars assessment for individual AF patients.

The current study is based on a single centre data. Multi-centre and multi-scanner studies are essential to validate the robustness and the generalisation of the proposed method. However, the possible domain shift among multi-centre and multi-scanner data will pose potential challenges for accurate segmentation. Therefore, in the future work, we will investigate feasible solutions to cope with the domain shift problems and tackle the multi-centre and multi-scanner data.

# ACKNOWLEDGEMENTS

This study was funded by the British Heart Foundation Project Grant (Project Number: PG/16/78/32402).